\documentclass[manuscript]{aastex}

\usepackage{rotating}
\usepackage{lscape}
\usepackage{pdflscape}
\usepackage{longtable}
\usepackage[usenames,dvips]{color}

\begin{document}

\title{Gamma-ray spectral properties of mature pulsars: a two-layer model}

\author{Wang Y.\altaffilmark{1}, Takata J.\altaffilmark{2} and Cheng K.S.\altaffilmark{3}}
\affil{Department of Physics, University of Hong Kong, Pokfulam Road, Hong Kong}

\altaffiltext{1}{yuwang@hku.hk}
\altaffiltext{2}{takata@hku.hk}
\altaffiltext{3}{hrspksc@hkucc.hku.hk}

\begin{abstract}
We use a simple two-layer outer gap model, whose accelerator
consists of a primary region and a screening region, to discuss  $\gamma$-ray
spectrum of mature pulsars detected by $Fermi$.
 By solving the Poisson equation with
an assumed simple step-function distributions of the charge density in
 these two regions,
the distribution of the electric field and the curvature radiation
 process of the accelerated particles can be calculated.
 In the our model, the
 properties of the phase-averaged spectrum
 can be completely specified by three gap parameters,
i.e. the fractional gap size in the outer magnetosphere,
the gap current in the primary region and the gap size ratio between
the primary region and the total gap size. We discuss how these
parameters affect the spectral properties. We argue that although the radiation
 mechanism in the outer gap is  curvature radiation process,
 the observed gamma-ray spectrum can substantially deviate from the
simple curvature spectrum because the overall spectrum
consists of two components, i.e. the primary region and screening region. In
some pulsars the radiation from the screening region is so strong that the
photon index from 100MeV to several GeV can be as flat as $\sim 2$.
 We show the fitting fractional gap thickness of the canonical pulsars
increases with the spin down age.  We find that the total gap current
is about 50~\% of the Goldreich-Julian value and the  thickness of the screening region is a few percent of the total gap thickness. We also find that the predicted $\gamma$-ray luminosity is less dependent on the spin down power ($L_{sd}$)  for the pulsars with
$L_{sd}\ga 10^{36}$~erg/s, while the $\gamma$-ray luminosity decreases
with the spin down power for the pulsars with $L_{sd}\la 10^{36}$~erg/s.
This relation may imply that the major gap closure mechanism is
 photon-photon
pair-creation process for the pulsars with  $L_{sd}\ga 10^{36}$~erg/s,
 while the magnetic pair-creation process for the pulsars with  $L_{sd}\la 10^{36}$~erg/s.
\end{abstract}

\section{Introduction}
The particle acceleration and high-energy $\gamma$-ray radiation
process in the pulsar
magnetosphere has been studied  with  polar cap  model
(Ruderman \& Sutherland  1975; Daugherty \& Harding 1982, 1996),
the slot gap model (Arons 1983; Muslimov \& Harding  2003; Harding
et al. 2008) and the outer gap model (Cheng, Ho \& Ruderman 1986a,b,
hereafter CHRa,b;
Hirotani 2008; Takata, Wang \& Cheng 2010).
All models have  assumed that the charged particles are accelerated by
the electric field along the magnetic field lines, and that the
acceleration region arises in the charge deficit region from the
Goldreich-Julian charge density (Goldreich \& Julian 1969).
The polar
cap model  assumes the acceleration region near the stellar
surface, while  the
outer gap model assumes a strong acceleration in outer magnetosphere.
The slot gap model proposes  that  the accelerating electric field near the
rim of the polar cap can not be completely screened out and the particles
are continuously accelerated up to high
altitude along the magnetic field lines.

The recent $\gamma$-ray instruments  have started to
break the bottleneck of study of the high-energy radiation from the
pulsar. In particular, the $Fermi$ LAT has detected $\gamma$-ray
emissions from 46~pulsars just in one year observation,
 including 21 radio-selected pulsars, 16 $\gamma$-ray selected pulsars,
the Geminga and 8 millisecond pulsars (Abdo et al. 2010a, 2009a,b),
while the pulsed radio emissions are reported from several
$\gamma$-ray selected pulsars (Camilo et. al. 2009). 
Romani \& Watters (2010) study the pulse profiles of canonical  pulsars observed by $Fermi$
LAT. They compute the pulse profiles predicted
 by the outer gap model and slot gap model with the magnetic field,
 which is composed of the vacuum dipole field plus the current-induced
field. They argued that the outer gap geometry is statistically more
  consistent with the observations than the slot gap geometry. On the
  other hand, the present population study  has argued that both
 outer gap and slot gap model  may not explain all features of Fermi
pulsars; for example, the both model predicts too few young pulsars
compared with the $Fermi$ pulsars (c.f.  Grenier \& Harding 2010). 
  Venter, Harding \& Guillemot (2009) fit the pulse profiles of the
 $Fermi$ detected millisecond pulsars with the geometries predicted by
the different emission models.  However, they
found that the pulse profiles  of  two out of
eight   millisecond pulsars cannot be fitted by either
 the  geometries with the outer gap or  the caustic models.  They proposed
 a pair-starved polar cap model, in which the multiplicity of the pairs
is not high enough to completely screen the electric field above the
polar cap,  and the particles are continuously accelerated up to high
altitude over the entire open field line region.

The observed spectra  also allow us to study the
site of the $\gamma$-ray emissions in the pulsar magnetosphere.  In the
first $Fermi$ pulsar catalog (Abdo et al. 2010a), the spectral fits have
been done assuming exponential cut-offs.
For example, the $Fermi$ LAT measured a detail spectral properties of
the Vela pulsar (Abdo et al. 2009c, 2010b). It is found that
the observed  spectrum is well fitted by a power
 law plus hyper-exponential
cutoff spectral model of the form $dN/dE\propto E^{-a} \exp[-(E/E_c)^b]$
with $b\le 1$ and $E_c\sim 1.5$~GeV.
 This implies that an  emission process in
 the outer magnetosphere is more favored than that near the stellar
 surface,  which predicts a super exponential
cut-off ($b>1$) due to the magnetic pair-creation prosses of
 the $\gamma$-rays.   Furthermore, the detection of the radiation above
25~GeV bands associated with the Crab pulsar by the MAGIC telescope
has also implied the
emission process in the outer magnetosphere (Aliu et al. 2008).

There is a common idea at least in different outer gap models.
Charged deficit regions, in which electric field
along magnetic field lines is not zero, exist in the outer
magnetospheric regions. The size of outer gap cannot
occupy the entire open field line region, because electrons/positrons
in the gap are accelerated to extremely relativistic energies and
each of these charged particles radiates a very large number of
GeV-photons due to the curvature of the magnetic field. Only a small
fraction of curvature photons converted
into pairs can limit the size of the outer gap
 (e.g. CHRa,b; Zhang \& Cheng 1997). Cheng, Ho \& Ruderman (1986a) have
 argued that a realistic dynamic gap
should consist of two regions, i.e. a primary acceleration  region of
lower part of the gap,
where seed current can produce multi-GeV gamma-ray photons,
 and a screening region of upper part of the gap,
 where the created electron/positron pairs
start to screen out the gap electric field (cf. Fig.12c of CHRa). More
detail numerical studies
(e.g. Hirotani 2006; Takata et al. 2006) expect that a screening region
must exist and in fact most gap current is located in this region.

In this paper we assume a simple two-layer gap structure, which
 consists of a primary acceleration region and a screening region. The electric
field along the magnetic field lines in both regions are not zero but
the exact distribution depends on the charge distribution in these
two regions. Charged particles accelerated in these regions radiate
 different curvature spectrum due to the position dependent electric field.
The combined spectrum depends sensitively on the charge distribution in
 these two acceleration regions. 
The first  purpose of this paper is that  we explain the photon indics of
 phase-averaged spectrum of mature pulsars by applying  our two-layer
 outer gap model. Observationally the phase-averaged spectrum above
 100~MeV bands  of
$Fermi$ $\gamma$-ray pulsars can be fit by a single power law plus
 exponential cut-off form (cf. Abdo et al. 2010a). It is found that
the photon index of the power law component
 distributes between $p\sim 0.5-2.5$ and the distribution has a peak
 around $p\sim 1.3$ (Abdo et al. 2010a).  It is virtually impossible to
 explain the very soft spectrum with a photon index $p>2$ with a simple
 curvature emission model by adjusting the curvature radius of a single
 layer. We will argue that  the two-layer model
can predict  the photon index as flat as $p\sim 2$ due to the overall
spectrum consists of the two components,
i.e. the curvature radiations from  screening region and from
the primary region.

The second purpose of this paper is to study the relation between
the $\gamma$-ray
 luminosity and the spin down luminosity.
 The increases of the $\gamma$-ray pulsars allow us
 to perform a more detail statistical study. For example, the results of
 the $Ferm$ will modify the relation between the $\gamma$-ray luminosity
($L_{\gamma}$) and the spin down power ($L_{sd}$),
for which $L_{\gamma}\propto L_{sd}^{\beta}$ with $\beta\sim 0.5$ were
 predicted by $EGRET$ $\gamma$-ray pulsars (Thompson 2004).  However,
 the luminosity $L_{\gamma}$ calculated from the observed flux include
 ambiguity due to the uncertainties of the distance to the pulsars and
 of the beaming factor. In this paper, therefore, we theoretically
predict $\gamma$-ray luminosity by fitting the properties of the
 phase-averaged spectrum. This will be important to  know  the true
 relation between the $\gamma$-ray luminosity and the spin down power.

 This paper is organized as follows.
In section 2, we present a simple two dimensional gap model, which
ignores the gap structure in the azimuthal direction. By assuming simple
step function of the charge distributions in these two acceleration
region, we can solve the Poisson equation analytically and obtain the
electric field everywhere inside the gap. We calculate the curvature
radiation to compare the observed phase-averaged spectrum. In section 3
we discuss  how the model parameters affect the spectral properties. We
also show the fitting results for the phase-averaged spectrum of the all
$Fermi$ detected gamma-ray pulsars except the Crab-like pulsars,
whose radiation spectra could be synchrotron-self-Compton process instead of
curvature radiation process. In section 4 we discuss the implications of the fitted parameters and deduced parameters. We present a brief conclusion in section~5.

\section{Theoretical Model}
\subsection{Gap structure with two-layer}

In addition to pulsar parameters, i.e. rotation period and dipole magnetic field, there are many factors to affect the observed detailed
spectrum of individual pulsar, e.g. the phase-resolved spectrum,
which depends on the magnetic field structure, the gap structure,
the inclination angle, viewing angle etc. However the phase-averaged
spectrum is less sensitive to these factors. In fact it can tell us
the most crucial factors, which govern the general  properties in
phase-averaged spectra of all pulsars. It is the aim of this paper to use
a simple  gap model to describe the acceleration and radiation in the outer gap, and try to identify the key parameters to
affect the spectral properties of pulsars. We consider the outer gap
accelerator in the magnetic meridian, which includes the rotation axis
and the magnetic axis. We divide the outer gap accelerator into
two-layers, namely the main acceleration region and the screening
region,  in the trans-field direction in the
magnetic meridian.
Fig.~\ref{structure} illustrates our gap structure with two-layer.

We express the electric potential in  the gap as
 $\Phi=\Phi'+\Phi_{0}$, where $\Phi_0$ is the co-rotating
potential, which satisfies $\bigtriangledown^2\Phi_0=-4\pi\rho_{GJ}$
 with $\rho_{GJ}$ being  the Goldreich-Julain charge density.  In addition,
the potential $\Phi\rq{}$ is the so called non co-rotating   potential,
 which  represents the deviation from co-rotating  potential and
 generates the accelerating electric field. Using the Poisson equation
 $\bigtriangledown^2\Phi=-4\pi\rho$, and  assuming the derivative of the potential field in the azimuthal direction is much smaller than those  in the poloidal plane, we express the Poisson equation of $\Phi'$ in the simple 2-dimensional geometry  as
\begin{equation}
\left(\frac{\partial^2}{\partial x^2}
+\frac{\partial^2}{\partial z^2}\right)\Phi'=-4\pi(\rho-\rho_{GJ}),
\label{Poisson0}
\end{equation}
where the coordinates $x$ and $z$ are distance along and perpendicular to
the magnetic field lines, respectively. Moreover, we assume that (1) the
 derivative  of the potential field in the trans-field (z) direction
is larger than that along (x-direction) the  magnetic field line and (2)
the variation of the Goldreich-Julian charge density in the trans-field
direction is negligible compared with that  along the magnetic field
line.  Although these approximations will be crude  for the pulsars that
have the gap size in the trans-field direction comparable to the size of
the magnetosphere,  we may apply these approximations for all pulsars
because the typical strength of the electric field in the gap might be
described by the solution of the first order approximation.
 By ignoring the derivative of the potential field
along the magnetic field line, we reduce the two-dimensional Poisson
equation~(\ref{Poisson0}) to one-dimensional form that
\begin{equation}
 \frac{\partial^2{}}{\partial{z^2}}\Phi'(x,z)=-4\pi [\rho(x,z)-\rho_{GJ}(x)],
\label{Poisson1}
\end{equation}
where we approximately express the Goldreich-Julain charge density as $\rho_{GJ}(x)\approx -\frac{\Omega Bx}{2\pi cs}$ (cf. Eqs. 3.2 and 3.3 of CHRa), where $\Omega$ is the angular velocity,
$B$ is the magnetic field strength,
 $s$ is the curvature radius of the magnetic field lines and $s\sim
 R_{lc}$ with $R_{lc}=c/\Omega$ being the light radius.

To solve the equation~(\ref{Poisson1}) we impose the boundary conditions
on the lower and upper boundaries. In this study, we consider that the
lower boundary is located on the last-open field line, and the upper
boundary is located on a fixed magnetic field line.
For the lower boundary, the condition that
\begin{equation}
\Phi'(x, z=0)=0
\end{equation}
is imposed.  On the upper boundary, we impose the gap closure conditions that
\begin{equation}
\Phi'(x, z=h_2)=0~~ \mathrm{and}~~
E'_{\perp}(x, z=h_2)=0,
\label{cond1}
\end{equation}
where $E'_{\perp}(x,z)=\partial \Phi'/\partial z$, and $h_2(x)$ is
the gap thickness measured from the last-open field line. The gap closure conditions  ensure that the total potential
(co-rotational potential +
non co-rotational potential) field in the gap is continuously connected
to the co-rotational potential field outside the gap.
We note that  three boundary conditions are imposed for
 the two order differential equation, implying the location of
the upper boundary can not be arbitrary chosen,
because both  Dirichlet-type and  Neumann-type conditions are
 imposed on the upper  boundary. In fact, the condition
$E'_{\perp}(x, z=h_2)=0$ connects the charge density distribution
and the gap thickness $h_2(x)$

As illustrated in Fig.~\ref{structure}, we divide the gap into two
 regions, i.e. the primary
acceleration region and the screening region.
The main acceleration region is  expanding between the last-open
 field line and the height $z=h_1(x)$ (region I in Fig.~\ref{structure}a),
and the screening region is expanding  between $z=h_1(x)$
and $z=h_2(x)$ (region II in Fig.~\ref{structure}a).
We note that
the averaged charge density in the outer gap accelerator
 must be less than the Goldreich-Julian  value, that is
$|\rho|<|\rho_{GJ}|$, in the primary acceleration region to arise a
 strong electric field along the magnetic field line.
In the screening region,
on the other hand,   $|\rho|>|\rho_{GJ}|$
is required to satisfy both boundary conditions in equation~(\ref{cond1}).
Physically speaking, the charge deficit in the primary region
is compensated by the charge excess in the screening region.

As demonstrated by
the electrodynamic studies (e.g. figure 7 in Hirotani 2006),
 (1) the charge density rapidly increases in the trans-field direction
at the height  where the  pair-creation process becomes
 important, and (2) the increase of the charge density
 will be saturated around upper boundary because of
 the screening of the electric field due to the pairs.  In this paper,
therefore, we describe the  distribution of the charge density
 in the trans-field direction (z-direction) as a step function form
(Fig.~\ref{structure}b);
\begin{equation}
\rho(x,z)=\left\{
   \begin{array}{ccc}
            \rho_1(x), &$if$& 0\leq z\leq h_1(x),\\
            \rho_2(x), &$if$& h_1(x)<z\leq h_2(x).
   \end{array}\right.
\end{equation}
We define the boundary $h_1(x)$ with a magnetic line as well as
$h_2(x)$, implying $h_2(x)/h_1(x)$ is nearly  constant along the
magnetic field line.

Using the boundary conditions that $\Phi'_z(z=0)=0$ and
 $\Phi'_z(z=h_2)=0$ and imposing the continuity of the potential field  $\Phi'_z$ and  $\partial
 \Phi'_z/\partial z$  at the height $h_1$,
 we obtain the solution of the Poisson equation~(\ref{Poisson1})
 as
\begin{equation}
\Phi'(x,z)=-2\pi \left\{
   \begin{array}{ccc}
            \{\rho_1(x)-\rho_{GJ}(x)\}z^2+C_1z, &$for$& 0\leq z\leq
	     h_1(x)\\
\{\rho_2(x)-\rho_{GJ}(x)\}
(z^2-h_2^2(x))+D_1(z-h_2(x)), &$for$& h_1(x)\leq z\leq h_2(x)
   \end{array}\right.
\label{potential}
\end{equation}
where
\[
 C_1(x)=\frac{(\rho_1-\rho_{GJ})h_1(h_1-2h_2)-(\rho_2-\rho_{GJ})(h_1-h_2)^2}
{h_2}
\]
and
\[
 D_2(x)=\frac{(\rho_1-\rho_{2})h_1^2-(\rho_2-\rho_{GJ})h_2^2}
{h_2}
\]

The gap closure condition  $E'_{\perp}=0$ on the upper
 boundary provides the relation among
 the charge densities $(\rho_1,~\rho_2)$ and the
gap thickness $(h_1,~h_2)$ that
\begin{equation}
[\rho_2(x)-\rho_{GJ}(x)]h^2_2(x)+(\rho_1(x)-\rho_2(x))h_1(x)^2=0.
\label{condi}
\end{equation}
During the fitting process (section~\ref{fitting}), we will find that
the charge density in the main
acceleration region is much smaller than that in the screening region,
that is, $|\rho_1|<<|\rho_2|$. This implies that the typical charge
density in the screening region is describes as
$\rho_2(x)\sim \rho_{GJ}(x)/[1-(h_1/h_2)^2]$. Namely, both screening
conditions $\Phi'(z=h_2)=0$ and $E'_{\perp}(z=h_2)=0$ are satisfied when
the charge density in the screening region is proportional to the
Goldreich-Julian charge density.  In the realistic situation, therefore,
we expect that the typical charge density in the screening region is
proportional to the Goldreich-Julian charge density.  For the charge
density, $\rho_1$, in the main acceleration region, we also put the
charge density proportional to the Goldreich-Julian charge density
for simplicity, $\rho_1\propto \rho_{GJ}$, because (1)  it is expected that the
charge density increases along the magnetic field region due to the
photon-photon pair-creation process and (2) in fact, the distribution of the
charge density in the main acceleration region is
 less important to the Potential field in the gap
(c.f. equation~(\ref{potential})) as long as the charge density
$|\rho_1|$ is much smaller than $|\rho_2|$ and $|\rho_{GJ}|$. We will see
in section~\ref{fitting},
 the observed spectrum can be fitted with $\rho_1\sim 0.1\rho_{GJ}$ and
$\rho_2\ge \rho_{GJ}$.  Because  we consider the
case that inclination angle is smaller than $\alpha<90^{\circ}$ and because the
outer gap extends beyond the null charge surface,  we rewrite the
charge density as
\begin{equation}
\frac{\rho-\rho_{GJ}}{\rho_{GJ}}=\left\{
   \begin{array}{ccc}
            -g_1, &$if$& 0\leq{}z\leq{}h_1\\
            g_2, &$if$& h_1<z\leq{}h_2
   \end{array},\right.
\end{equation}
where $g_1>0$,  $g_2>0$ (cf. Fig.~\ref{structure}c).   The accelerating
electric field along the magnetic field line, $E'_{||}=-\partial
\Phi'/\partial x$ is described as
\begin{equation}
E'_{||}(z)\sim\frac{\Omega B }{c s}\left\{
   \begin{array}{ccc}
            -g_1z^2+C'_1z, &$for$& 0\leq z\leq
	     h_1(x)\\
g_2
(z^2-h_2^2(x))+D'_1(z-h_2(x)), &$for$& h_1(x)\leq z\leq h_2(x)
   \end{array}\right.
\label{electric}
\end{equation}
where
\[
 C'_1(x)=-\frac{g_1h_1(h_1-2h_2)+g_2(h_1-h_2)^2}
{h_2}
\]
and
\[
 D'_2(x)=-\frac{(g_1+g_2)h_1^2+g_2h_2^2}
{h_2},
\]
where we used $\partial (Bh_2^2)/\partial x\sim 0$, $\partial
(z/h_2)/\partial x\sim 0$, $\partial (h_1/h_2)/\partial x\sim 0$ and
$\partial s/\partial x\sim0$.

 A particular solution of equation~(\ref{potential})  is obtained by
specifying any three parameters or their combinations out of four,
i.e. $h_1$, $h_2$, $g_1$ and $g_2$.
In the next section, we fit the observed phase-averaged spectra with
the following combination of model parameters,
i.e. $h_2(R_{lc})/R_{lc}$, $1-g_1$  and $h_1/h_2$, which correspond
to the fractional gap size, the physical gap charge density (or current)
 divided by Goldreich-Julian value
 in the primary region and the gap size ratio between the primary and
screening regions.


\subsection{Curvature radiation spectrum}
The charged particles are accelerated by the electric field along the
magnetic field lines in the gap,
 and  emit  $\gamma$-rays via the curvature radiation.
 The  Lorentz factor, $\gamma_e$, can be estimated from the condition that
$eE'_{\parallel}(z)c=l_{cur}(z)$, where
$l_{cur}(z)=2e^2c\gamma^4_e(z)/3s^2$ is  the power of the curvature
 radiation, as
\begin{equation}
\gamma_e(z)=\left[\frac{3}{2}\frac{s^2}{e}E'_{\parallel}(z)\right]^{1/4}.
\label{lorent}
\end{equation}
For a single particle, the spectrum of curvature radiation is described by
\begin{equation}
F_{cur}^{single}=\frac{\sqrt{3}e^2\gamma_e}{2\pi{}\hbar{}sE_{\gamma}}
F(E_{\gamma}/E_{cur})
\end{equation}
where $E_{cur}(z)=(3/2)\hbar{}c\gamma_e^3(z)/s$
is the characteristic energy of the radiated curvature photon
and $F(x)=\int^{\infty}_{x}K_{5/3}(y)dy$, where $K_{5/3}$ is the
modified Bessel functions of order 5/3.
The total curvature radiation spectrum from the outer gap  is computed
from
\begin{equation}
F_{cur}(E_{\gamma})=\int n(x,z) F_{cur}^{single}(z)dV,
\end{equation}
where $n(x,z)\sim n_{GJ}(x)(1+g(z))$  is the number density, $n_{GJ}(x)=\Omega
B(x)/2\pi c e$,  and $dV=\delta A_{\perp}(x) dx$ is the volume of the
element, where
 $\delta A_{\perp}(x)$ is the increment area of width $\delta z$.
The increment are is calculated as
 $\delta A_{\perp}(x)\sim 2\pi R_{lc}\delta z(R_{lc})
B(R_{lc})/B(x)$, where $\delta z(R_{lc})$ is the
width at the light cylinder and
we used the magnetic flux conservation
that $B(x)\delta A_{\perp}(x)=B(R_{lc})\delta A_{\perp} (R_{lc})\sim
2\pi B_{lc}R_{lc}\delta z(R_{lc})$.

 The total flux received at Earth is
 \begin{equation}
 F(E_{\gamma})=\frac{1}{\Delta{\Omega}d^2}F_{cur}(E_{\gamma}),
 \end{equation}
where $d$ is the distance of the pulsar and  $\Delta{\Omega}$ is the solid angle.

\section{Results}
\subsection{Properties of the curvature spectra with the gap structures}
\label{property}
Basically we have three independent fitting parameters, i.e the
fractional gap size defined by $f\equiv h_2(R_{lc})/R_{lc}$,
the number density (current) in the primary region ($1-g_1$) and the ratio between the thicknesses of  primary region and total gap size $h_1/h_2$.
Qualitatively the fractional gap size mainly determines  the total potential drop in the gap and the strength of the accelerating field,  and hence it affects the intensity and the cut-off energy of the curvature radiation. We will see that the number density ($1-g_1$) in the primary region determines the slope of the curvature spectrum below the cut-off energy  and the ratio ($h_1/h_2$) controls the spectral break in lower energy bands around 100~MeV. In below we give the quantitative description  how these three parameters affect the spectrum.

Fig.~\ref{frac} summarizes the dependency of the curvature spectra on the
gap thickness. The lines present  the spectra with  the
fractional gap thickness of $f$=0.8~(solid line), 0.6~(dashed line), 0.4~(dotted line) and 0.2~(dashed-dotted line). The results are for $1-g_1=0.10$, $h_1/h_2=0.933$ and the parameters of the Geminga pulsar.
We can see in Fig.~\ref{frac} that  the spectral properties are
sensitive to the gap thickness. Specifically, the intensity increases
with the gap fractional thickness. This is because  the total potential
drop in the gap and therefore
 magnitude of the accelerating electric field are  proportional to the
 square of the gap thickness ($\Phi'\propto f^2$), as
equations~(\ref{potential}) and~(\ref{electric}) imply. Therefore,
the predicted  flux of the curvature radiation from the gap is
approximately proportional to $E_{\gamma}^2F\propto \Phi'\times N\propto
f^3$, where we used that the total number of the particles (or current)
in  the gap is proportional to $N\propto f$.  Fig.~\ref{frac} also shows
that the spectrum becomes hard with
increase of the fractional gap thickness.  From the
equations~(\ref{electric}) and~(\ref{lorent}), we can see that  the
typical energy of the curvature photons is proportional to
  $E_{\gamma}\propto (3/2)\hbar c\gamma_e^3/s\propto f^{3/2}$, implying
the cut-off energy in spectrum increases with the fractional gap
thickness as Fig.~\ref{frac} shows.

Fig.~\ref{g1} shows how the charge density, $\rho=(1-g_1)\rho_{GJ}$,
 of the primary region affects the curvature
spectrum. The lines represent the curvature spectra
for $1-g_1=0.30$ (solid line), $0.10$ (dashed line), 0.05 (dotted line)
and $0.01$~(dashed-dotted line). The results are for $f=0.76$, $h_1/h_2=0.933$
and the parameters of the Geminga pulsar are used. Qualitatively speaking, the number density in the primary region determines the  slope of the spectrum below  the cut-off energy. We  can find in Fig.~\ref{g1} (e.g. dashed-dotted line) that the spectrum is  divided into two
components, that is,  higher and lower energy components.
The higher
energy component ($E_{\gamma}\ga 1$~GeV) comes from
the primary region,  where the charge density is smaller than
the Goldreich-Julain charge density,
while  the other one ($E_{\gamma}\la
1$~GeV) originates  from the screening  region, where the charge density
exceeds the Goldreich-Julain charge density. Because the accelerating
electric field in the primary region is stronger  than that in the
screening region, the Lorentz factor of the particles and the resultant
typical energy of the curvature photons are larger
in the primary region than in the screening
region.

 The screening condition, $(h_2/h_1)^2=1+g_1/g_2$
(c.f. Eq.\ref{condi}), implies a larger number density (corresponding to larger $g_2$) in the screening region is required for a smaller number density (corresponding to  larger $g_1$)
in the primary region.
In such a case,
if the number density of the primary region is much smaller than Goldreich-Julian value,  the energy flux of the curvature radiation from the primary region becomes
small compared with that from the screening region, as indicated by the dashed-dotted line in Fig~\ref{g1}.
 With the present model, therefore,   number density in the primary region,  and resultant  ratio of the total numbers of particles in the primary and the screening regions  mainly determines
the slope of the spectrum below the cut-off energy.
Fig.~\ref{g1} also shows that the cut-off energy in  the spectra in
GeV energy bands decreases as  the charge  density in the primary region
(or $1-g_1$) increases.
This is because the deviation from the Goldreich-Julian
charge density becomes small as the charge  density increases, implying  that the accelerating electric field and the resultant typical
energy of the curvature photons decrease with increasing of the charge
density.

Fig.\ref{h1_h2} summarizes  the dependency of the properties  of the curvature
spectrum on the ratio of the thicknesses
between the primary region and the screening region, $h_1/h_2$.
The lines represent the curvature spectra
for $h_1/h_2=0.70$ (solid line), $0.80$ (dashed line), 0.90 (dotted line)
and $0.96$~(dashed-dotted line) with  $f=0.76$, $1-g_1=0.1$
and the parameters of the Geminga pulsar.  We note that the increase
of the ratio $h_1/h_2$ corresponds to  the decrease of the thickness
of the screening region ($h_2-h_1$) relative to the primary region.  In the present model, the ratio $h_1/h_2$ determines the energy width between the spectral cut-off in GeV band and the spectral break in 100~MeV bands, because the curvature spectrum of the  primary region has an energy peak at several GeV, while that of the screening region shows an energy peak at several hundred MeV.   In the dashed-dotted line in Fig.~\ref{h1_h2} ($h_1/h_2=0.96$), for example, the spectral cut-off appearing at several GeV is a result of  the emissions from the primary region, while the spectral break seen  around 200~MeV is produced  by the emissions from the screening region. As the ratio decreases, on the other hand, the position of the spectral break in lower energy  shifts to higher energy (e.g. $\sim$500~MeV for $h_1/h_2=0.90$), and the energy width between the cut-off  energy  and the spectral break  energy  decreases.  This is because the thickness of the screening region ($h_2-h_1$) relative to that of the primary region ($h_1$) becomes thick as the ratio decreases, implying
 the flux of the emissions from the screening region increases  and its spectrum becomes hard.  For the ratio $h_1/h_2=0.70$ (solid line in Fig.~\ref{h1_h2}),
the emissions from the screening region dominates the emissions from the primary region.

\subsection{Fitting Results}
\label{fitting}
We fit the phase-averaged spectra of the 42 $\gamma$-ray pulsars measured by the $EGRET$ and
$Fermi$ telescopes with the present model. For the observed spectra,
 we use information reported in  the first $Fermi$ catalogue (Abdo et al. 2010a), in which the observed  data were fit with a single power law  plus exponential cut-off form. In our  fitting, we do not include the Crab-like  young
pulsars (e.g. the Crab pulsar, PSR J1124-5916)
because the radiation mechanism of the Crab-like pulsars are synchrotron-self-Compton process instead of curvature
radiation process (Cheng, Ruderman \& Zhang 2000; Takata \& Chang 2007).
In the Crab-like pulsars their soft photon density is sufficiently high so that most curvature photons from the outer gap will be converted into pairs within the light cylinder and the observed gamma-rays resulting from
the inverse Compton scattering of the synchrotron photons of secondary pairs. This radiation process differs from what
we have considered in this paper and therefore we will not consider them.

 As discussed in section~\ref{property}, our fitting parameters are the fractional gap thickness $f$, the charge density in the primary region, $1-g_1$, and the ratio $h_1/h_2$. In fact, given observed phase-averaged spectrum
 can be uniquely fit with one set of $(f, 1-g_1, h_1/h_2)$ with a small uncertainty. The model fitting was proceeded as following. First, we deduced  the typical fractional gap thickness, $f$, from  the observed intensity and the cut-off energy, because the fractional gap thickness $f$  greatly affects the intensity and the cut-off energy as Fig.~\ref{frac} shows. For the next step,  we fit the spectral slope below the cut-off energy with  the charge density in the primary region, $1-g_1$, which mainly controls  the slope of the calculated spectrum as Fig.~\ref{g1} shows.  Finally, we determined  $h_1/h_2$, which controls the spectral  width  between the cut-off energy in several GeV and the spectral break energy in lower energy bands.

Fig.~\ref{canonical} presents  the fitting results with the observed data for the 6 canonical pulsars.  The data points are taken from  the EGRET observations (Fierro, 1995) for the Geminga, PSRs J057-5226,J1709-4229 and  J1952+3252,
and from the $Fermi$ observations for the Vela (Abdo et al. 2009c and
2010b) and  PSR J2021+4206 (Trep et al. 2010). The grey strips represent
the errors of the photon index, cut-off energy and intensity measured by
the $Fermi$ observations. The solid lines represent  the best fit
spectra with the fitting parameters  listed in each panel and in Table~1.
We use the dashed and dashed-dotted lines in the panels of the Gemiga and Vela
pulsars to preset  how the best fit parameters include the
uncertainties. The dotted lines are results for the fractional gap
thickness ($f=0.7$ for the Geminga and 0.145 for the Vela) about 10~\%
difference than the best fitting values, while the dashed-dotted lines are results for
$1-g_1$ (0.13 for the Geminga and 0.1 for the Vela) and
$h_1/h_2$ (0.867 for the Geminga and 0.967 for the Vela) about 10-20~\%
difference than the best fitting values. It is obvious that both dashed and
dashed-dotted lines can not explain the observed data, implying the
fitting parameters include uncertainties of about 10~\% for the Geminga
 and Vela pulsars.

Figs~\ref{canofit} and \ref{milifit} compare  fitting results  with the $Fermi$ observations for the 28 canonical pulsars and 8 millisecond pulsars, respectively.  The solid line and dashed line are corresponding to  the model spectra and the observations, respectively. The grey strips represent the errors of the observations.  In Figs~\ref{canofit}-\ref{milifit}, we can see that the present model can reproduce well  the observations,  although there is  a small discrepancy between the model and the observations around 100~MeV energy bands for some pulsars (such as PSRs J1907+06 and J2229+6144).

In Table~1, we summarize  the observed pulsar parameters (second-fifth columns)
 and the fitting parameters (sixth-ninth columns).  Fig.~\ref{ftau} plots  the  fitted fractional gap thickness as a function of the spin down age $\tau$. We find in Fig.~\ref{ftau} that  the fitted fractional gap size $f_{fit}$ (sixth column) is between 0.1-0.9 and is
bigger than the estimated fractional gap size of the Crab pulsar (e.g. CHRb estimated $f_{Crab} \sim 0.05$).  We can   see a clear  trend  in Fig.~\ref{ftau} that  the fractional gap thickness for the canonical pulsars  increases with the spin down age.
This is expected because the potential drop of the pulsar decreases as the spin down age increases, implying a larger fractional gap thickness is required to accelerate the particles to emit several GeV $\gamma$-ray photons via the curvature radiation process.  For the millisecond pulsars, it is required more number of sample to argue the relation between the fractional gap size and the spin down age.

In seventh column of Table 1  we can see that the current in the primary region is roughly $\sim 10\%$ of the Goldreich-Julian current. This is important
because if the gap current in this region is too large then  accelerating electric field is not strong enough to accelerate the
gap electrons/positrons to sufficiently high Lorentz. Consequently multi-GeV photons cannot be generated and pair creation process cannot take place.

The ratio between the primary region and the total gap size $h_1/h_2$ is less than but very closed to unity. This indicates that the screening region is very thin and is only a few percents of the total gap size. This is expected because although the mean free path ($\lambda$) of photon-photon pair-creation  is longer than the light cylinder radius in the case of mature pulsars, the thickness of the screening region is substantially reduced by the photon multiplicity ($N_{\gamma}$) and it can be estimated as
\begin{equation}
h_2-h_1 \approx \frac{\lambda^2}{sN_{\gamma}^2}
\end{equation}
(cf. equation 5.7 and C21 of CHRa). Taking $\lambda \sim 10^3 R_{lc}$, $s\sim R_{lc}\sim
10^9 cm$ and $N_{\gamma}\sim 10^4$, $h_2(R_{lc})-h_1(R_{lc}) \sim
10^7~\mathrm{cm}<<h_2(R_{lc})\sim 10^8-10^{9}$cm. Also the magnetic
pair-creation near the stellar surface could supply the screening pairs
within a thin layer (Takata et al. 2010).

 In order to match the observed
gamma-ray flux, we need to introduce one more fitting parameter,
i.e. $\Delta \Omega d^2$ (ninth column), which is the product of gamma-ray solid angle
and the distance to the pulsar. Using the observed distance (fourth column) and the fitting $\Delta \Omega d^2$, we estimate the solid angle
$\Delta \Omega_{fit}$ (twelfth column).  We can see that the averaged
value  of the deduced solid angle $\Delta \Omega_{fit}$  is order of
unity, which is usually assumed. On the other hand, some $\gamma$-ray
pulsars are required a solid angle  much larger ($>10$) or smaller
($<0.1$) solid angle than the averaged value.  This would reflect the effects of
 the viewing geometry. The fitting solid angle much larger (or smaller)
 than the mean value may imply that the pulsars are observed with a
 viewing angle which cuts through the emission region where the
 intensity is much smaller (or larger) than the mean intensity.

For the observed distance to pulsar, we can see that first of all the
error of the observed distance is usually quite large and secondly there
are some pulsars without the measured distance. Therefore if the solid
angle of gamma-ray pulsars is the same, which  is usually assumed to be
unity, then we can use the theoretical predicted power and the observed
flux to  predict the distance to the pulsar,
which is  listed in the last column of Table 1.

\section{Discussion}
In Table~1 we deduce the averaged gap current in units of the Goldreich-Julain value (tenth column). In addition to show that the curvature radiation emitted by the accelerated charged particles can consistently explain the observed spectrum
of mature pulsars detected by $Fermi$, it is interesting to note that
from Table~1 that the gap current is roughly constant. Specifically, the
variation of the gap current is less than 20\% and the gap current is
around 50-60\% of the full Goldreich-Julian current.
Although it is not easily explain why the total current in the gap
is close  to half of Goldreich-Julian value,
this result appears
qualitatively reasonable because if the current is very closed to the
Goldreich-Julian current, the electric field will be substantially
reduced. Consequently the charged particles cannot be accelerated to
extremely relativistic to emit multi-GeV photons and hence pair creation
process cannot occur. In fact we speculate that the gap may operate in a
dynamical format, namely, the gap can begin with an almost charge
starvation state, and then charged particles are accelerated to
extremely relativistic and emit $10^4- 10^5$ photons per particle across
the gap. A small fraction of these photons are converted into pairs,
which can almost immediately quench the gap,  and the current in this
stage is almost full Goldreich-Julian current. Since particles and
photons are all moving in speed of lights, the on/off time scales of
these two stages should be very similar and which may be  order of
$R_{lc}/c$. After averaging over time, it may not be surprised to have a
time average gap current
about 50\% of the Goldreich Julian value. If there were good time
resolution, this speculation predicts that there is a time variability
in the $\gamma$-ray intensity with a time scale of $R_{lc}/c$. This may
be an evidence of the pair-creation process in the pulsar magnetosphere,
 which  maintains the global electric circuit of the star.

The predicted  $\gamma$-ray luminosity ($L^{fit}_{\gamma}$),  for each pulsar
is listed in eleventh column of Table~1, where we use
$L^{fit}_{\gamma}=f^3_{fit}L_{sd}$ with
$L_{sd}=(2\pi)^4B_s^{2}R_s^6/6c^3P^4$ being 
the spin down luminosity.  We note that  uncertainties of
the  predicted $\gamma$-ray luminosity is small because  the
uncertainties of the fitting parameters is small, say  about 10~\%,
as discussed in section~\ref{fitting}. In Fig.~\ref{Lsd},
the predicted $\gamma$-ray luminosity is plotted as a function of
the spin down power $L_{sd}$.
We can see a trend in Fig.~\ref{Lsd} that the predicted $\gamma$-ray
luminosity
shows less dependency  on the spin down power for the pulsars with  $L_{sd}\ga 10^{36}$~erg/s, while $L_{\gamma}$ decreases with the spin down power for the pulsars with $L_{sd}\la 10^{36}$~erg/s.  This change of the dependency of the $\gamma$-ray luminosity on the spin down power may be caused by switching the gap closure process. In fact,  the solid and dashed lines in Fig.~\ref{Lsd} represent the relation predicted  by the outer gap  model with different gap closure processes as follows.

 Zhang \& Cheng (1997) have argued that when the pulsars cool down, the cooling X-rays may not be sufficient to convert
the curvature photons emitted by the accelerated particles into pairs to restrict the growth of the outer gap. They suggest that however the X-rays emitted by the heated polar cap due to the return current can provide a self-consistent mechanism to restrict the fractional size of the gap as
\begin{equation}
f_{ZC}=0.32P_{-1}^{26/21}B_{12}^{-4/7},
\end{equation}
where $P_{-1}$ is the rotation period in units of 0.1s.
This model predicts the $\gamma$-ray luminosity is related with the spin
down luminosity as $L_{\gamma}\propto L_{sd}^{1/14}B_{12}^{1/7}\sim
L_{sd}^{1/14}$, which is represented by the solid line in
Fig.\ref{Lsd}. This less dependency on the spin down luminosity  may
explain the behavior of the predicted $\gamma$-ray luminosity of  the
pulsars with $L_{sd}\ga 10^{36}$~erg/s.

Recently Takata et al. (2010) argue that since the electric field
decreases rapidly from the null charge surface to the inner boundary,
where the gap electric field must vanish, the radiation loss cannot be
compensated by the acceleration of the local electric field. When this
occurs the Lorentz factor of the incoming electrons/positrons is
determined by the equating the radiation loss time scale and the
particle crossing time scale. It is interesting to note that the
characteristic photon energy of curvature radiation is independent of
pulsar parameters and is given by $m_ec^2/\alpha_f\sim 100 \rm MeV$,
where $\alpha_f$ is the fine structure constant.

They further argue that these 100MeV photons can become pairs by magnetic
pair creation process. These secondary pairs can continue to radiate due
to synchrotron radiation and the characteristic energy of synchrotron
photons is of order of several MeV. The photon multiplicity is easily
over $10^5$ per each incoming particle. Such cascade process has also
been considered before (e.g. Cheng \& Zhang 1999). For a simple dipolar 
field structure, all these pairs
should move inward and they cannot affect the outer gap. However they
argue that the existence of strong surface local field (e.g. Ruderman
1991, Arons 1993) has been widely suggested. In particular if the field
lines near the surface, instead of nearly perpendicular to the surface,
are bending sideward due to the strong local field. The pairs created in
these local magnetic field lines can have an angle bigger than
90$^{\circ}$, which results in an outgoing flow of pairs. In fact it
only needs a very tiny fraction (1-10) out of $10^5$ photons creating
pairs in these field lines, which are sufficient to provide screening in
the outer gap when they migrate to the outer magnetosphere.

They estimate the
fractional gap thickness when this situation occurs as
\begin{equation}
f_m= 0.25 K   P_{-1}^{1/2},
\end{equation}
where  $K\propto B_{m,12}^{-2}s_7$ is the parameter to characterize the local parameters, e.g.
$B_{m,12}$ and $s_7$ are the local magnetic field in units of $10^{12}$G and the local curvature radius in units
of $10^7$cm. From this estimate they predict that the gamma-ray luminosity
 is related to the spin down power
 as $L_{\gamma} \propto L_{sd}^{5/8}$, which is represented by the dashed-line in Fig~\ref{Lsd}. This gap closure process  may explain the relation between the predicted  $\gamma$-ray luminosity  and the spin down power  for the pulsars with $L_{sd}\la 10^{36}$~erg/s, as  Fig.~\ref{Lsd} indicates.

\section{Conclusion}
\begin{figure}
\centering
\includegraphics[width=\textwidth]{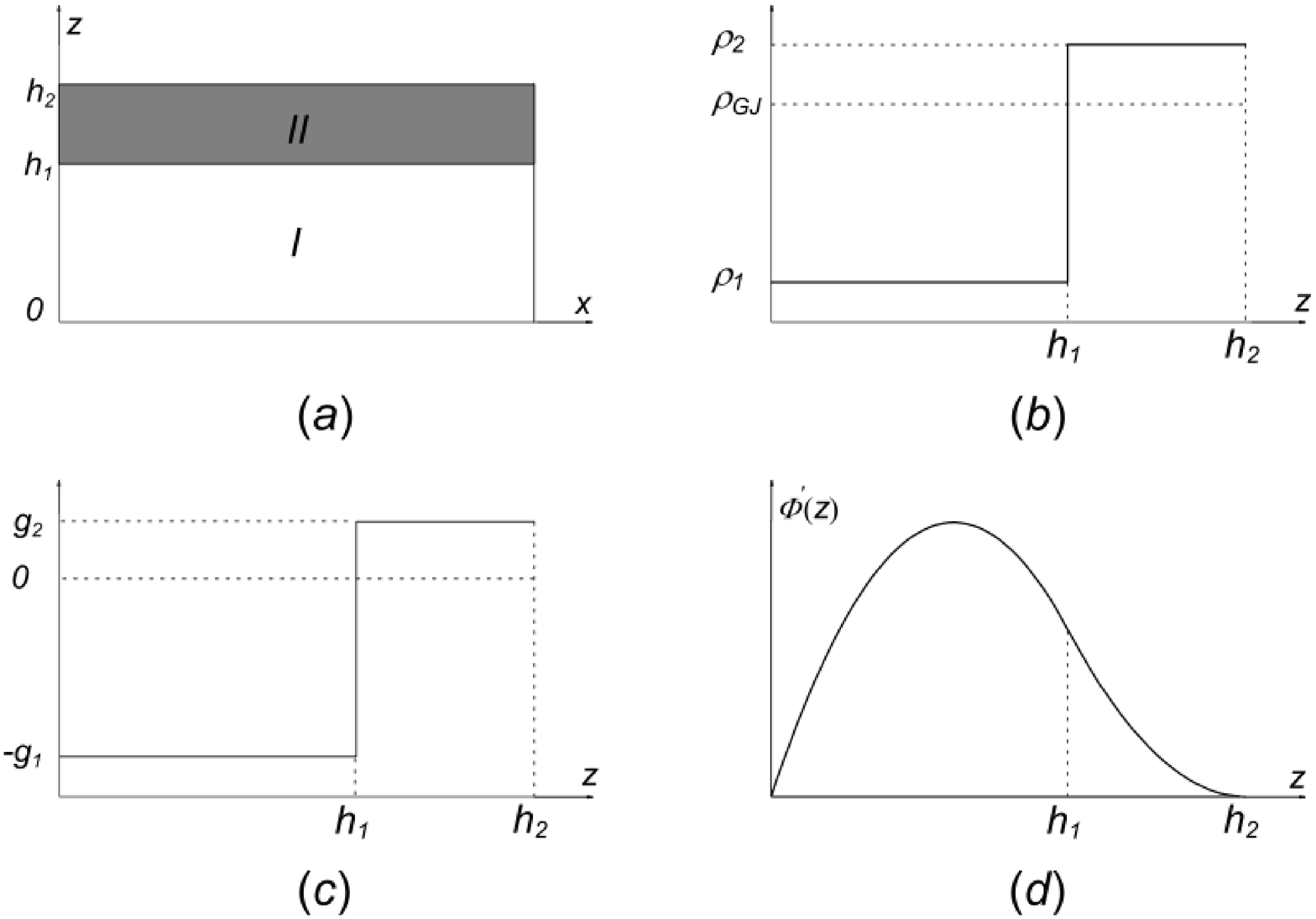}
\caption{Schematic view  of the gap structure. (a)The structure of the
 outer gap with two-layer. The regions $I$ and $II$ are the primary region and the screening region respectively. (b)The distribution of the charge density in the trans-field ($z$) direction in the gap. (c)The distribution of the $g(z)$ in $z$-direction in the gap. (d)The distribution of the $\Phi\rq{}(z)$, and the accelerating electric field $E'_{||}(z)\propto \Phi\rq{}(z)$.}
\label{structure}
\end{figure}

\begin{figure}
\centering
\includegraphics{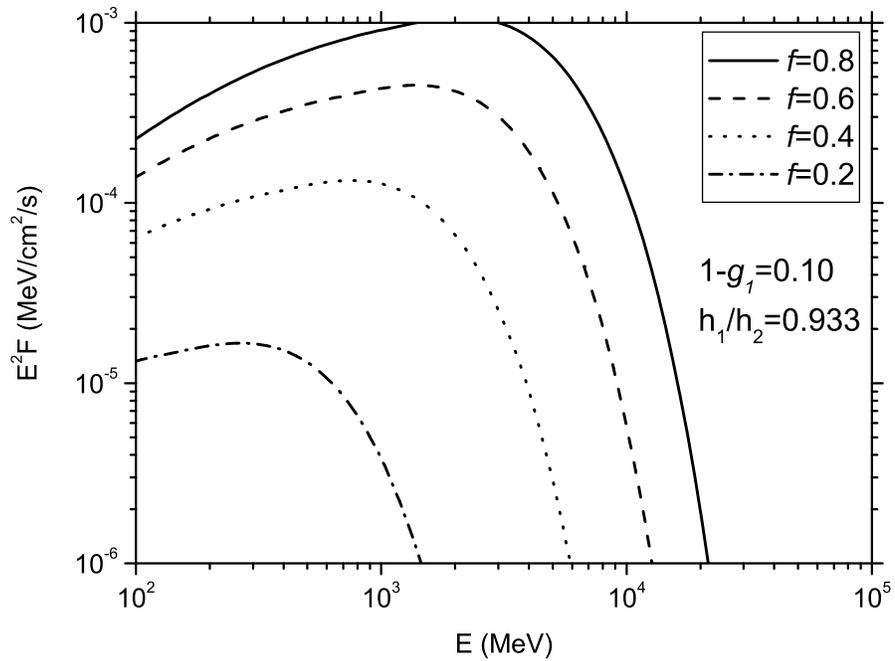}
\caption{The effects of $f$ on the shape of the spectrum of the curvature radiation. The results are for $1-g_1=0.1$ and $h_1/h_2=0.933$. The parameters of Geminga are used.}
\label{frac}
\end{figure}

\begin{figure}
\centering
\includegraphics{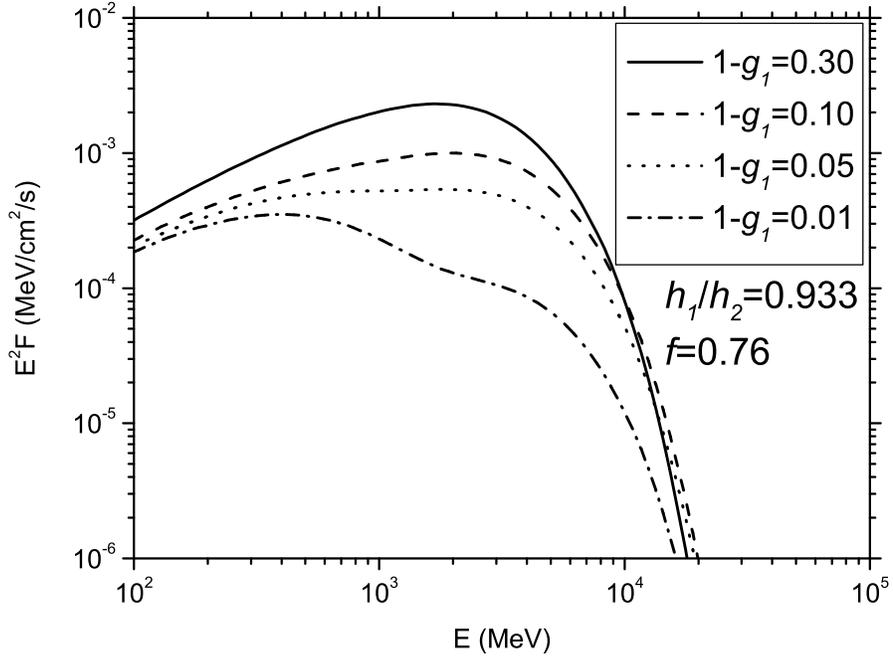}
\caption{The effect of the charge density of the main acceleration region, $1-g_1$, on the shape of the spectrum, where $h_1/h_2=0.933$ and $f=0.76$ are used. The parameters of Geminga are used.}
\label{g1}
\end{figure}

\begin{figure}
\centering
\includegraphics{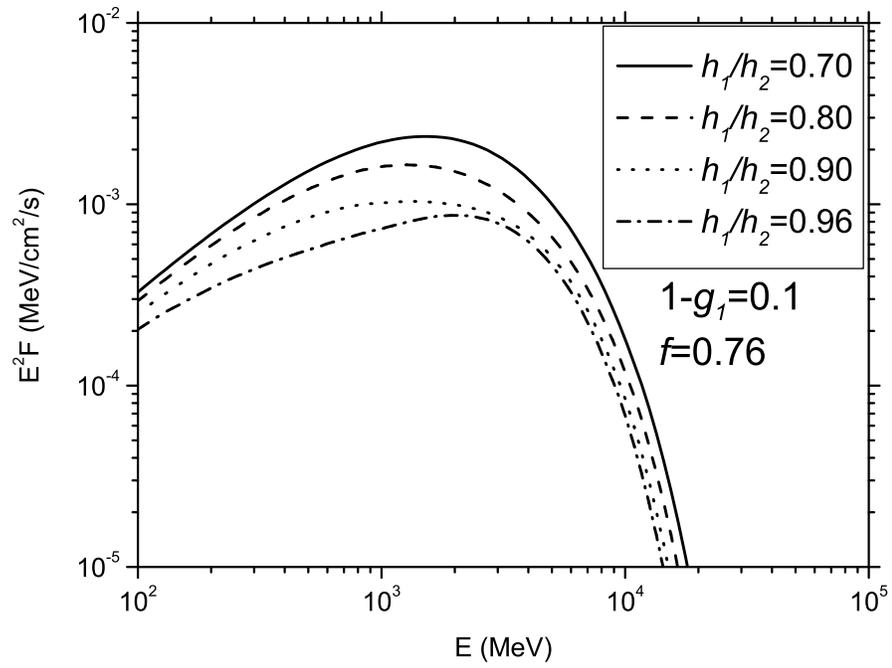}
\caption{The effect of $h_1/h_2$ on the shape of the spectrum with the $1-g_1=0.1$ and $f=0.76$. The parameters of Geminga
 are used.}
\label{h1_h2}
\end{figure}

\begin{figure}
\centering
\includegraphics[width=\textwidth]{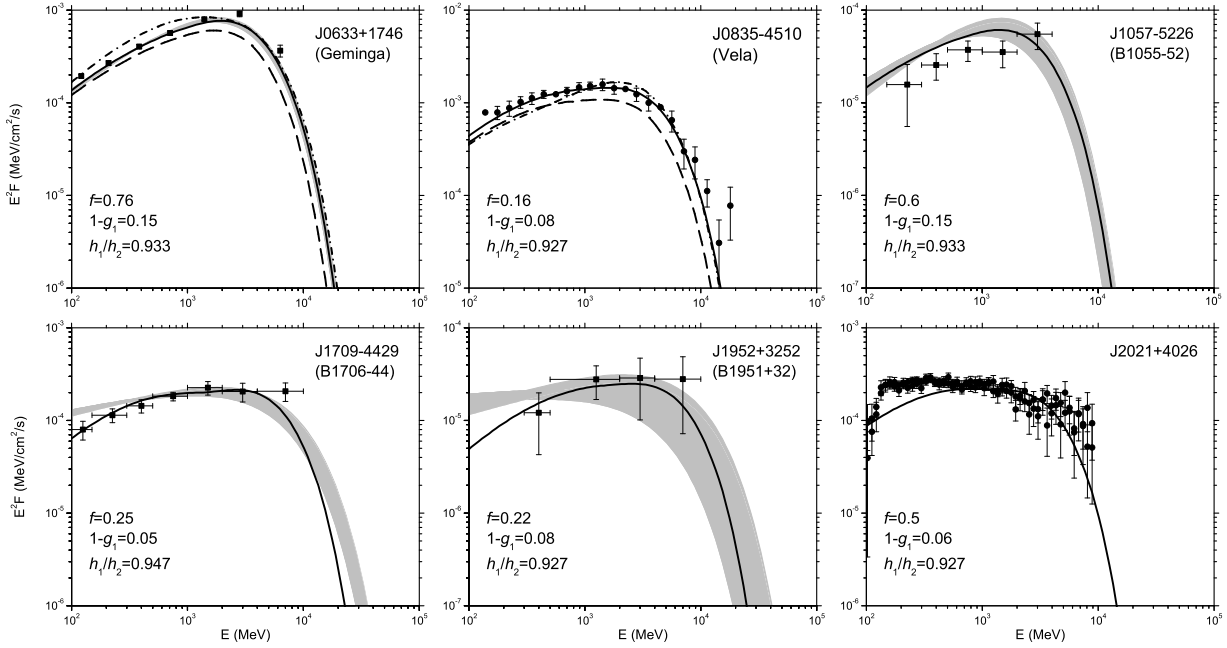}
\caption{The model results of 6 canonical pulsars. The solid lines represent the best fitting model spectra with the fitting parameters listed in each panel.  The circles are the observed data from the $Fermi$ LAT, which are taken from  Abdo et al. (2009d) for the Vela pulsar  and  Trep et al. (2010) for J2021+4026. The boxes are the observed data from $EGERT$ (Fierro 1995).
The dotted lines in the panels of the Geminga and the Vela pulsars are results for the fractional gap thickness ($f=0.7$ for the Geminga and 0.145 for the Vela) about 10~\% difference than the best fitting fractional gap thickness, while the dashed-dotted lines are results for $1-g_1$ (0.13 for the Geminga and 0.1 for the Vela) and  $h_1/h_2$ (0.867 for the Geminga and 0.967 for the Vela) about 10-20~\% difference than the best fitting parameters.
}
\label{canonical}
\end{figure}

In this paper, we applied the two-layer  outer gap model  to fit
the observed phase-averaged spectra of the 42~$\gamma$-ray pulsars
detected by the $Fermi$ telescope.  Our gap structure consists
 two parts, which are the primary acceleration region and the
screening region (Fig.~\ref{structure}). In the primary acceleration
region, the charge density is less than the Goldreich-Julian charge
density ($|\rho|<|\rho_{GJ}|$), while in the screening region, the
charge density exceeds  $\rho_{GJ}$ to screen out the accelerating
electric field.  Assuming  a step function of distribution of charge
density in the direction perpendicular to the magnetic field lines,
we solve the Poisson equation to obtain the accelerating electric field
in the gap. We fit the observed phase-averaged spectrum with the three
fitting parameters, that is, the fractional gap thickness $f$, the
number density of the primary region ($1-g_1$), and the ratio of the
thicknesses of the primary and the secondary regions, $h_1/h_2$.
 We demonstrated that the gap thickness affects the position of the
 cut-off energy (several GeV)  appeared in  the curvature spectrum, as
Fig.\ref{frac} shows.  We also showed that the number density of the
 primary region and the resultant ratio of the number densities between
 the primary and the screening region determine the slope of the
 spectrum below the cut-off energy (Fig.~\ref{g1}), and that the ratio
$h_1/h_2$ mainly affects the position of the spectral break  at  100~MeV
bands (Fig.\ref{h1_h2}), which is caused by the emissions from the
screening region.  With two-layer mode, the observed soft spectrum with
the photon index of $\sim 2$ for the some pulsars can be explained by
the overall  spectrum consists of the two components, i.e. the curvature
radiations from screening region and from the primary region. 
 Our fitting results show that the fractional gap
 thickness for the canonical pulsars  tends to increases with the spin
down age, as Fig~\ref{ftau}. shows. The observations can be fit by the
ratio $h_1/h_2$ smaller than but close to unity, implying the screening
region is  only a few percent of total gap thickness. The present model
predicts the gap current is about 50~\% of the Goldreich-Julain value.
We found that the predicted $\gamma$-ray luminosity shows less
dependency on the spin down power  for the pulsars with
$L_{sd}\ga 10^{36}$~erg/s, while it decreases with the spin down power
 for  pulsars with $L_{sd}\la 10^{36}$~erg/s (Fig.~\ref{Lsd}).
 We discussed the relation of the $\gamma$-ray luminosity and the
spin down power  with the gap closure mechanisms of Zhang\& Cheng
(1997), which predict $L_{\gamma}\propto
L_{sd}^{1/14}$  and of Takata et al. (2010), which predicts
$L_{\gamma}\propto L_{sd}^{5/8}$.

While the present simple model can be applied to   discuss the observed
 phase-averaged spectra, a three-dimensional model should be required to
discuss the observed light curves and the phase-resolved spectra.
The detail properties of the observed light curves
(e.g. number of the peak and phase of peak) and the phase-resolved
spectra will reflect  the three-dimensional distributions of the number
 density of particle and electric field in the gap.
 For example, the $Fermi$ telescope revealed the third peak, whose
 position depends on the energy bands,  in the light curve of the
Vela pulsar (Abdo et al. 2009d, 2010). The third peak  in GeV energy
 bands, which has not been expected by the previous emission models,
will be understood  with a detail  three-dimensional analysis.
 In the subsequent papers, therefore, we will extend the present
two-dimensional analysis into a three-dimensional one, and fit the
phase resolved spectra and the energy-dependent light curve for the
individual pulsar.

\acknowledgments
We wish to express our thanks to the referee for his/her insightful
comments on the manuscript. We thank the useful discussions with
H.-K.~Chang, K.~Hirotani, C.Y.~Hui, B.~Rudak, M.~Ruderman and
S.~Shibata.
 This work is supported by a GRF grant of Hong Kong SAR Government
 under HKU700908P.

\newpage

\begin{figure}
\centering
\includegraphics[width=\textwidth]{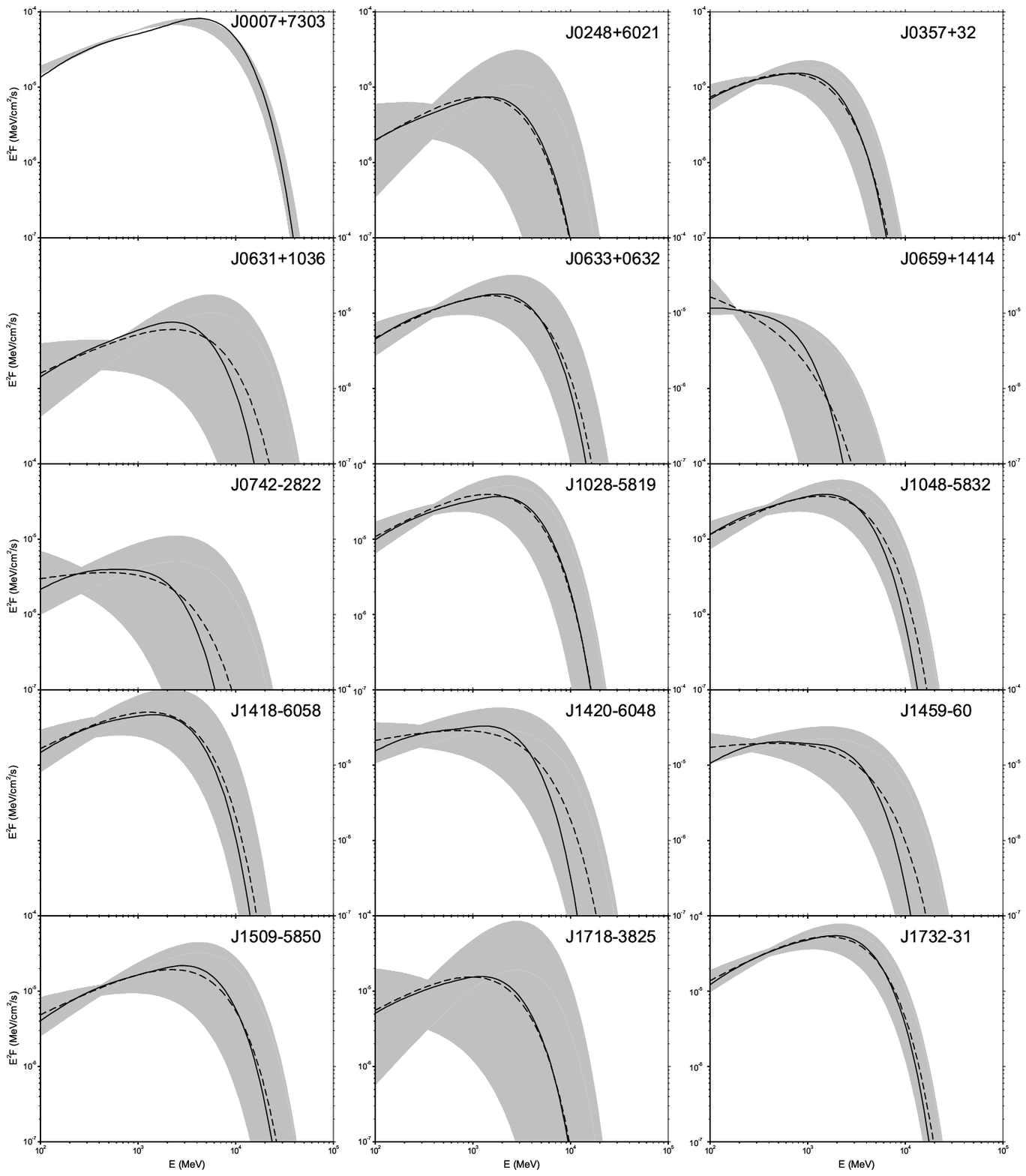}
\end{figure}
\begin{figure}
\centering
\includegraphics[width=\textwidth]{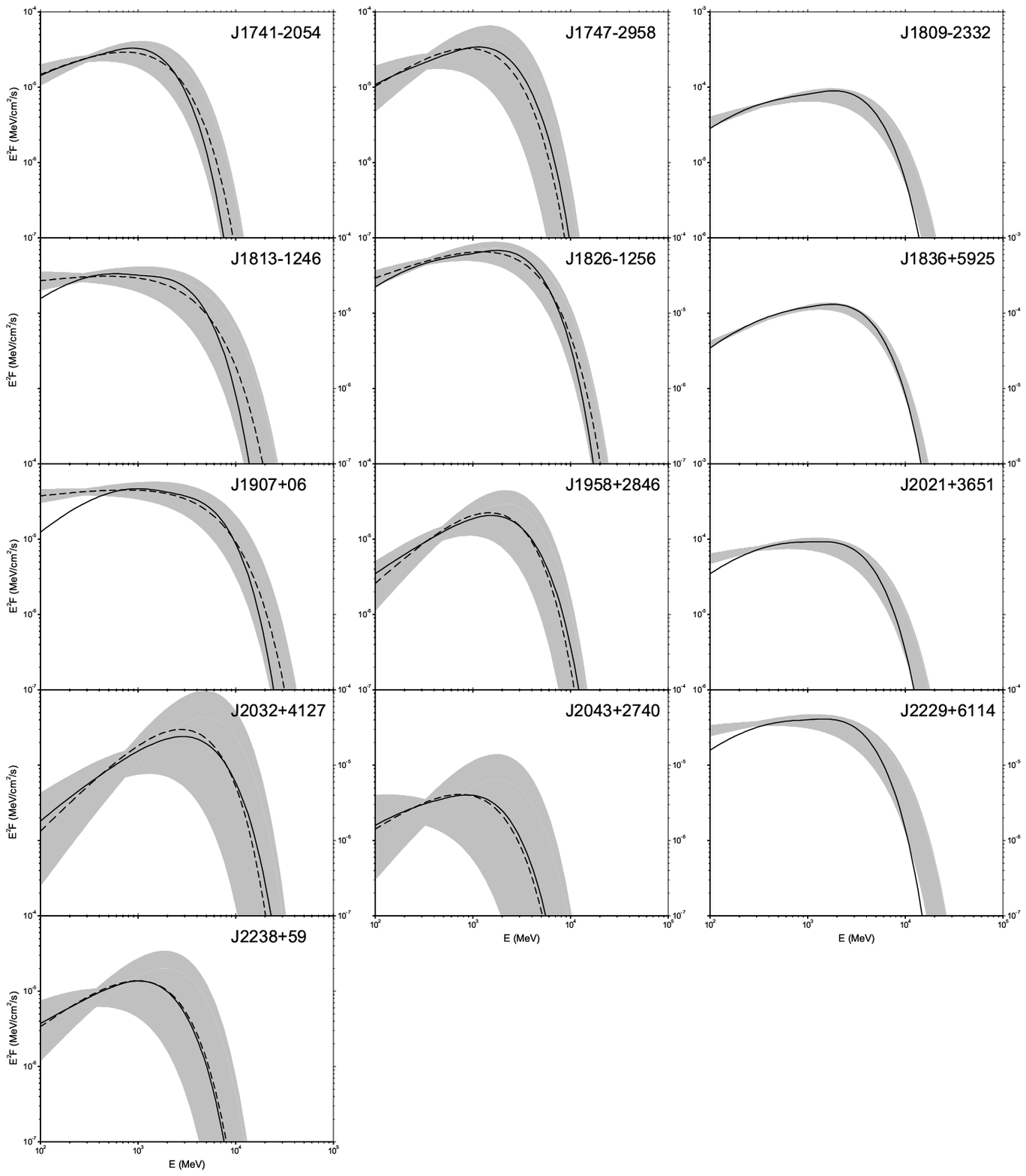}
\caption{The model results of 28 canonical pulsars. The solid lines are model results, the dash lines correspond to the photon indexes, cut off energies and photon fluxes from the $Fermi$ catalogue and the grey strips represent the errors of the three(Abdo et al. 2009a).}
\label{canofit}
\end{figure}

\begin{figure}
\centering
\includegraphics[width=\textwidth]{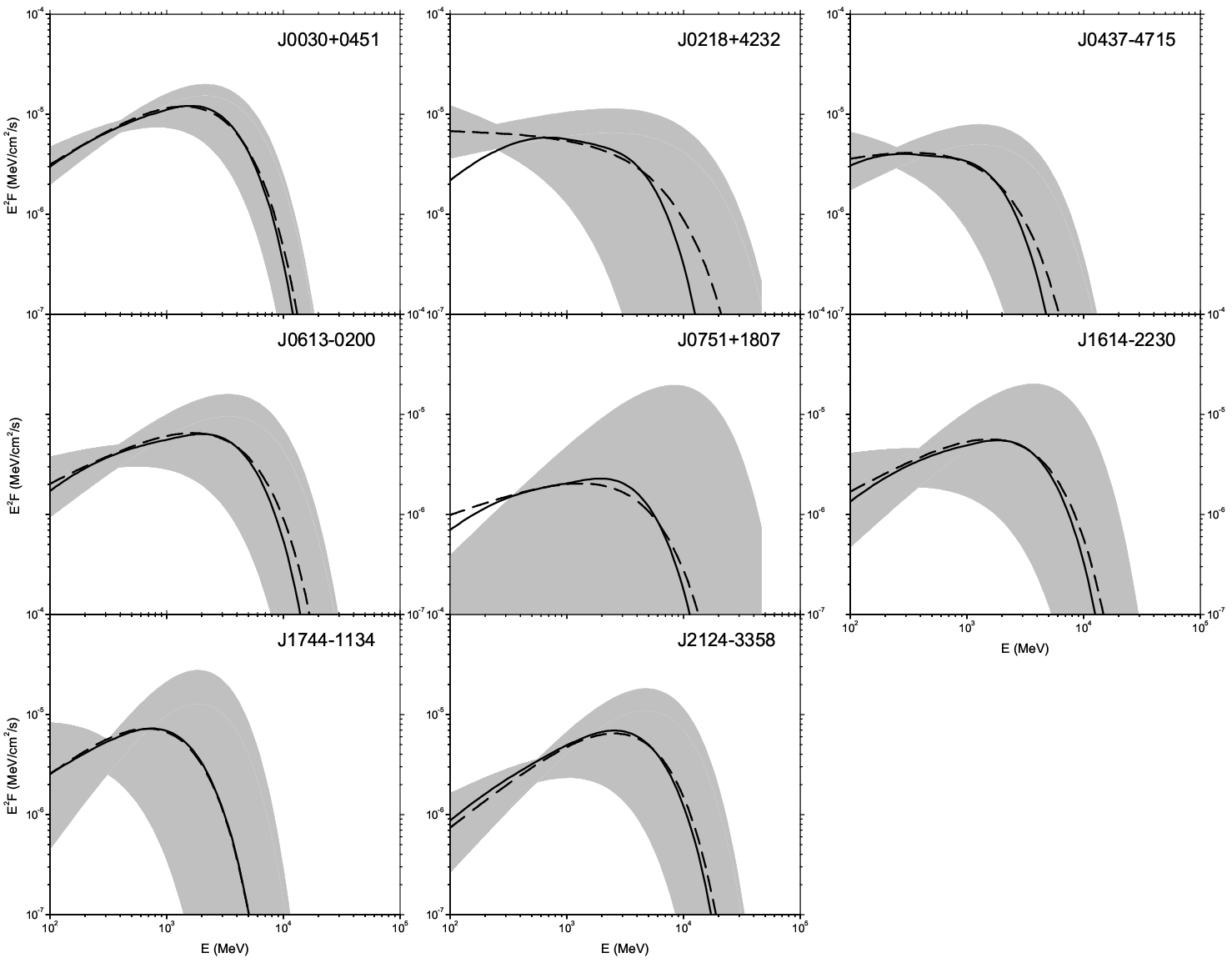}
\caption{The model results of the millisecond pulsars. The solid lines are model results,the dash lines correspond to the photon indexes, cut off energies and photon fluxes from the $Fermi$ catalogue and the grey strips represent the errors of the three(Abdo et al. 2009a).}
\label{milifit}
\end{figure}

\begin{figure}
\centering
\includegraphics[width=\textwidth]{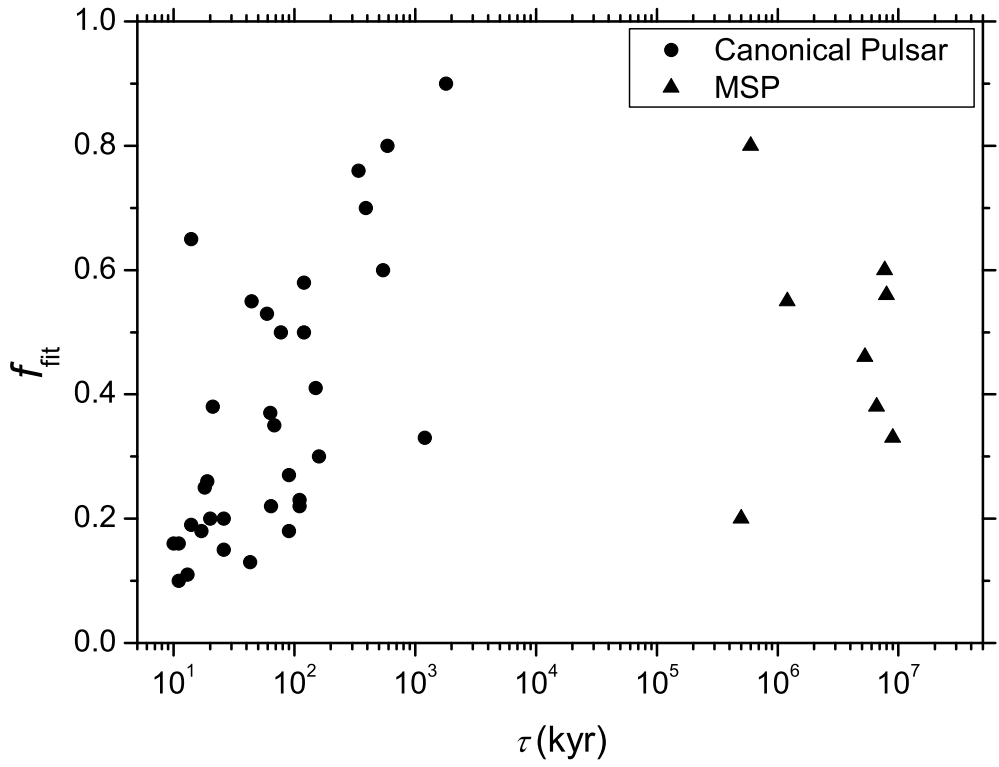}
\label{ftau}
\caption{The fitting fractional gap thickness as a function of the spin down age. The circle and triangle represent for the canonical and millisecond pulsars, respectively. }
\end{figure}

\begin{figure}
\centering
\includegraphics[width=\textwidth]{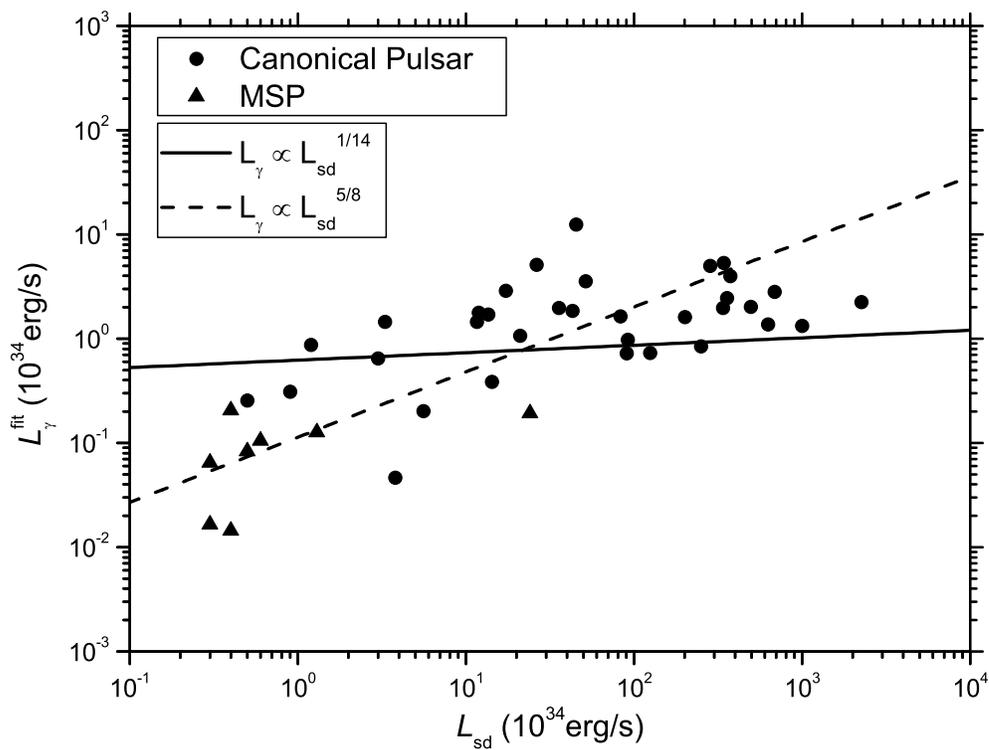}
\caption{The predicted $\gamma$-ray luminosity as a function of the spin down power. The circle and triangle represent for the canonical and millisecond pulsars, respectively. The solid and dashed lines represent the models by Zhang \& Cheng (1997) and Takata et al. (2010), respectively. }
\label{Lsd}
\end{figure}

\begin{landscape}
\scriptsize

\begin{longtable}{lllcccccccccl}

\caption{Parameters}
\endfirsthead

\caption{continued}
\endhead


\hline\hline
\multicolumn{1}{c}{} &
\multicolumn{4}{c}{Observed Parameters} &
\multicolumn{4}{c}{Fitting Parameters} &
\multicolumn{4}{c}{Deduced Parameters}\\

 Name & $P$ (ms) & $B_{12}$ & $d^{obs}$ (kpc) & $F_{100}^{obs}$(10$^{-8}$ph cm$^{-2}$s$^{-1}$)& $f_{fit}$ &
  $1-g_1$ & $h_1/h_2$ & $\Delta{\Omega}d^2$ (kpc$^2$) & $\eta_{gap}$ & $L_{\gamma}^{fit}$ (10$^{33}$erg/s) & $\Delta{\Omega}_{fit}$ & $d(\Delta{\Omega}=1)$\\
 \hline
 J0007+7303$^*$ &316&10.6&$1.4\pm0.3$&$30.7\pm1.3$&
 0.65&0.06&0.967&4.508&0.538&
 124.1&$2.3^{+1.42}_{-0.74}$&2.12\\ 

 J0248+602 &217&3.44&2-9&$3.7\pm1.8$&
 0.37
 & 0.10&0.953&6.875&0.561&
 10.63&0.08-1.72&2.62\\

 J0357+32$^*$ & 444& 1.9 & $\cdots$ & $10.4\pm1.2$ &
 0.80
 &0.12&0.927&0.72&0.577&
 2.56& $\cdots$ & 0.85\\

 J0631+1036 &288&5.44&0.75-3.62&$2.8\pm1.2$&
 0.55
 &0.10&0.953&18&0.561&
 28.78&1.37-32&4.24\\

 J0633+0632$^*$ & 297 & 4.84 & $\cdots$ & $8.4\pm1.4$ &
 0.53
 &0.10&0.947&4.81&0.562&
 17.72& $\cdots$ & 2.19\\

 J0633+1746 &237&1.59&$0.250^{+0.120}_{-0.062}$&$305.3\pm3.5$&
 0.76
 &0.15&0.933&0.125&0.590&
 14.49&$2^{+1.54}_{-1.09}$&0.35\\

 J0659+1414 &385&4.34&$0.288^{+0.033}_{-0.027}$&$10\pm1.4$&
 0.23
 &0.05&0.920&0.12442&0.545&
 0.4624&$1.5^{+0.32}_{-0.29}$&0.35\\

 J0742-2822 &167&1.67&$2.07^{+1.38}_{-1.07}$&$3.18\pm1.2$&
 0.30&0.08&0.920&
 4.2849&0.559&
 3.861&$1^{+3.28}_{-0.64}$&2.07\\

 J0835-4510 &89.3&3.40&$0.287^{+0.019}_{-0.017}$&$1061\pm7.0$&
 0.16&0.08&0.927&
 0.08237&0.557&
 28.18&$1^{+0.13}_{-0.12}$&0.29\\

 J1028-5819 &91.4&1.21&$2.33\pm0.70$&$19.6\pm3.1$&
 0.27&0.09&0.947&
 1.9544&0.557&
16.38&$0.36^{+0.38}_{-0.15}$&1.40\\

 J1048-5832 &124&3.48&$2.71\pm0.81$&$19.7\pm3.0$&
 0.20&0.10&0.947&
 1.98291&0.562&
 16.08&$0.27^{+0.28}_{-0.11}$&1.41\\

 J1057-5226 &197&1.08&$0.72\pm0.2$&$30.45\pm1.7$&
 0.60&0.15&0.933&
 0.72576&0.590&
 6.48&$1.4^{+1.28}_{-0.54}$&0.85\\

 J1418-6058$^*$ &111&4.37&2-5&$27.7\pm8.3$&
 0.16&0.10&0.940&
 2.2&0.564&
20.27&0.09-0.55&1.48\\

 J1420-6048 &68.2&2.38&$5.6\pm1.7$&$24.2\pm7.9$&
 0.11&0.06&0.947&
 1.2544&0.543&
13.31&$0.04^{+0.04}_{-0.02}$&1.12\\

 J1459-60$^*$ & 103 & 1.6 & $\cdots$ & $17.8\pm3.4$ &
 0.22&0.05&0.927&
 1.45&0.543&
9.786& $\cdots$ & 1.20\\

 J1509-5850 &88.9&0.90&$2.6\pm0.8$&$8.7\pm1.4$&
 0.41&0.09&0.960&
 7.098&0.554&
35.49&$1.05^{+1.14}_{-0.44}$&2.66\\

 J1709-4429 &102&3.04&1.4-3.6&$149.8\pm4.1$&
 0.25&0.05&0.947&
 0.63&0.538&
53.28&0.05-0.32&0.79\\

 J1718-3825 &74.7&0.99&$3.82\pm1.15$&$9.1\pm5.8$&
 0.18&0.11&0.947&
 2.48071&0.567&
7.29&$0.17^{+0.18}_{-0.07}$&1.58\\

 J1732-31$^*$ & 197 & 2.24 & $\cdots$ & $25.3\pm3.0$ &
 0.50&0.11&0.933&
 1.62&0.570&
17& $\cdots$ &1.27\\

 J1741-2054$^*$ &414&2.31&$0.38\pm0.11$&$20.3\pm2.0$&
 0.70&0.10&0.960&
 0.361&0.559&
3.087&$2.5^{+2.45}_{-1.00}$&0.60\\

 J1747-2958 &98.8&2.46&2-5&$18.2\pm4.2$&
 0.15&0.10&0.953&
 1.2&0.561&
8.471&0.05-0.30&1.10\\

 J1809-2332$^*$ &147&2.24&$1.7\pm1.0$&$49.5\pm3.0$&
 0.35&0.07&0.947&
 0.7225&0.548&
18.44&$0.25^{+1.22}_{-0.15}$&0.85\\

 J1813-1246$^*$ & 48.1 & 0.92 & $\cdots$ & $28.1\pm3.5$ &
 0.13&0.05&0.927&
 1.25&0.543&
13.75& $\cdots$ & 1.12\\

 J1826-1256$^*$ & 110 & 3.64 & $\cdots$ & $41.8\pm4.1$ &
 0.19&0.07&0.947&
 1.28&0.548&
24.56& $\cdots$ & 1.13\\
 \hline
 \newpage

\hline\hline
\multicolumn{1}{c}{} &
\multicolumn{4}{c}{Observed Parameters} &
\multicolumn{4}{c}{Fitting Parameters} &
\multicolumn{4}{c}{Deduced Parameters}\\

 Name & $P$ (ms) & $B_{12}$ & $d^{obs}$ (kpc) & $F_{100}^{obs}$(10$^{-8}$ph cm$^{-2}$s$^{-1}$)&$f_{fit}$ &
  $1-g_1$ & $h_1/h_2$ & $\Delta{\Omega}d^2$ (kpc$^2$) & $\eta_{gap}$ &$ L_{\gamma}^{fit}$ (10$^{33}$erg/s) & $\Delta{\Omega}_{fit}$ & $d(\Delta{\Omega}=1)$\\
 \hline

 J1836+5925$^*$ &173&0.51&$<0.8$&$65.6\pm1.8$&
 0.90
 &0.10&0.940&0.32&0.564&
8.748&$>0.5$&0.56\\

 J1907+06$^*$ & 107 & 3.09 & $\cdots$ & $40.25\pm3.8$ &
 0.26
 &0.05&0.920&3.81&0.545&
49.92& $\cdots$ & 1.95\\

 J1952+3252 &39.5&0.48&$2.0\pm0.5$&$17.6\pm1.9$&
 0.22
 &0.08&0.927&6.6&0.558&
39.82&$1.65^{+1.28}_{-0.59}$&2.57\\

 J1958+2846$^*$ & 290 & 7.95 & $\cdots$ & $7.65\pm1.6$ &
 0.38
 &0.20&0.967&8&0.607&
19.64& $\cdots$ & 2.83\\

 J2021+3651 &104&3.18&$2.1^{+2.1}_{-1.0}$&$67.35\pm4.4$&
 0.18
 &0.07&0.927&0.7938&0.553&
19.71&$0.18^{+0.48}_{-0.14}$&0.89\\

 J2021+4026$^*$ &265&3.84&$1.5\pm0.45$&$152.6\pm4.9$&
 0.50
 &0.06&0.927&0.28125&0.548&
14.5&$0.125^{+0.13}_{-0.05}$&0.53\\

 J2032+4127$^*$ &143&1.68&1.6-3.6&$6\pm2.3$&
 0.58
 &0.30&0.953&25.2&0.658&
51.31&1.94-9.84&5.02\\

 J2043+2740 &96.1&0.35&$1.80\pm0.54$&$2.41\pm0.90$&
 0.33
 &0.12&0.933&2.961&0.575&
2.012&$0.9^{+0.94}_{-0.37}$&1.71\\

 J2229+6114 &51.6&2.00&0.8-6.5&$32.6\pm2.2$&
 0.10
 &0.07&0.940&2&0.549&
22.5&0.05-3.125&1.41\\

 J2238+59$^*$ & 163 & 4.04 & $\cdots$ & $6.8\pm1.8$ &
 0.20
 &0.15&0.967&3.64&0.582&
7.224& $\cdots$ & 1.91\\
\hline

Millisecond pulsars\\
\hline \hline
\multicolumn{1}{c}{} &
\multicolumn{4}{c}{Observed Parameters} &
\multicolumn{4}{c}{Fitting Parameters} &
\multicolumn{4}{c}{Deduced Parameters}\\

 Name & $P$ (ms) & $B_{8}$ & $d^{obs}$ (kpc) & $F_{100}^{obs}$(10$^{-8}$ph cm$^{-2}$s$^{-1}$)&$f_{fit}$ &
  $1-g_1$ & $h_1/h_2$ & $\Delta{\Omega}d^2$ (kpc$^2$) & $\eta_{gap}$ & $L_{\gamma}^{fit}$ (10$^{33}$erg/s) & $\Delta{\Omega}_{fit}$ & $d(\Delta{\Omega}=1)$\\
 \hline
 J0030+0451 & 4.9 & 2.27 & $0.3\pm{}0.09$ & $5.83\pm{}0.78$ &
0.60
 &0.12 & 0.947 & 0.36& 0.572&
0.648&$4^{+4.16}_{-1.63}$&0.6\\ 

 J0218+4232 & 2.3& 4.14 & 2.5-4 &$6.2\pm1.7$&
0.20
&0.05 & 0.920 & 1.17& 0.545&
1.92 &0.07312-0.1872&1.08\\

 J0437-4715 & 5.8& 2.9 & $0.1563\pm{}0.0013$ & $3.65\pm{}0.84$ &
0.38
&0.05 & 0.927 & 0.12215& 0.543&
0.165 & $5^{+0.08}_{-0.08}$&0.35\\

 J0613-0200 & 3.1& 1.76 & $0.48^{+0.19}_{-0.11}$ & $3.38\pm0.85$ &
0.46
&0.08 & 0.947 & 0.8064& 0.553&
1.265 &$3.5^{+2.39}_{-1.70}$&0.898\\

 J0751+1807 & 3.5& 1.5 & $0.6^{+0.6}_{-0.2}$ &$1.35\pm0.66$ &
 0.56
 &0.07 & 0.947 & 1.62& 0.548&
 1.054 &$4.5^{+5.63}_{-3.38}$&1.272\\

 J1614-2230 & 3.2& 1.2 & $1.27\pm{}0.39$ & $2.89\pm1.2$ &
 0.55
 &0.10 & 0.933 & 0.80645& 0.566&
 0.832 &$0.5^{+0.54}_{-0.21}$&0.898\\

 J1744-1134 & 4.1& 1.8 & $0.357^{+0.043}_{-0.035}$ & $4.3\pm1.6$ &
 0.33
 &0.15 & 0.953 & 0.15294& 0.585&
 0.1438 &$1.2^{+0.28}_{-0.24}$&0.391\\

 J2124-3358 & 4.9& 2.4 & $0.25^{+0.25}_{-0.08}$ & $1.95\pm0.49$ &
 0.80
 &0.15 & 0.953 & 2& 0.585&
 2.048 &$32^{+37.2}_{-24}$&1.414\\
 \hline

\end{longtable}

\normalsize
Note. --- The first column is the name of the pulsar, the pulsar with
 "$^*$" means it is a $\gamma$-ray selected  pulsar, which
 was detected by $Fermi$ blind search. The second to fifth
column are observed parameters: periods, surface magnetic fields in
 units of $10^{12}$G, observed distances and the photon flux
 respectively. The data of these four columns are from the pulsar
 catalogue of the $Fermi$ LAT(Abdo et al. 2010a). $g_1$, $h_1/h2$,
 $f_{fit}(R_{lc})$ and $\Delta{\Omega}d^2$ are fitting parameters.
$\eta_{gap}\equiv\frac{\rho}{\rho_{GJ}}=\frac{(1-g_1)h_1
+(1+g_2)(h_2-h_1)}{h_2}$ is the average gap current in units of
 Goldreich-Julian current. $L_{\gamma}^{fit}=f_{fit}^3L_{sd}$ is the
 $\gamma$-ray luminosity of the model result. $\Delta{\Omega}_{fit}
\equiv\Delta{\Omega}d^2/d_{obs}^2$, the errors are due to the errors of
 the distances. The $d(\Delta{\Omega=1})$ is the predicted distance in
 units of kpc, it is obtained from the fitting parameter
 $\Delta{\Omega}d^2$, when $\Delta{\Omega}=1$.

\end{landscape}


\begin{thebibliography}{}
\bibitem[\protect\citeauthoryear{Abdo}{2010a}]{ab10a}
Abdo A.A. et al., 2010a, ApJS, 187, 460
\bibitem[\protect\citeauthoryear{Abdo}{2010b}]{ab10b}
Abdo A.A. et al., 2010b, ApJ, 713, 154
\bibitem[\protect\citeauthoryear{Abdo}{2009}]{ab09b}
Abdo A.A. et al., 2009a, Sci., 325, 848
\bibitem[\protect\citeauthoryear{Abdo}{2009}]{ab09c}
Abdo A.A. et al., 2009b, Sci., 325, 840
\bibitem[\protect\citeauthoryear{Abdo}{2009}]{ab09d}
Abdo A.A. et al., 2009c, ApJ, 696, 1084
\bibitem[\protect\citeauthoryear{Aliu}{2009}]{al09}
Aliu E. et al., 2008, Sci, 322, 1221
\bibitem[\protect\citeauthoryear{Arons}{1993}]{ar93}
Arons J., 1993, ApJ, 408, 160
\bibitem[\protect\citeauthoryear{Arons}{1983}]{ar83}
Arons J., 1983, ApJ, 266, 215
\bibitem[\protect\citeauthoryear{Camilo}{2009}]{ca09}
Camilo F. et al., 2009 ApJ, 705, 1
\bibitem[\protect\citeauthoryear{Cheng}{1986a}]{ch86a}
Cheng K.S., Ho C. \& Ruderman M. 1986a, ApJ, 300, 500
\bibitem[\protect\citeauthoryear{Cheng}{1986b}]{ch86b}
Cheng K.S., Ho C. \& Ruderman M. 1986b, ApJ, 300, 522
\bibitem[\protect\citeauthoryear{Cheng}{2000}]{ch00}
Cheng K.S., Ruderman, M \& Zhang L., 2000, ApJ, 537, 964
\bibitem[\protect\citeauthoryear{Cheng}{1999}]{ch99}
Cheng K.S. \& Zhang L., 1999, ApJ, 515, 337
\bibitem[\protect\citeauthoryear{Daugherty}{1982}]{da82}
Daugherty J.K. \& Harding A.K., 1982, ApJ, 252, 337
\bibitem[\protect\citeauthoryear{Daugherty}{1996}]{da96}
Daugherty J.K. \& Harding A.K., 1996, ApJ, 458, 278
\bibitem[\protect\citeauthoryear{Fierro}{1995}]{fi95}
Fierro J.M., 1995, PhD thesis, Stanford Univ
\bibitem[\protect\citeauthoryear{Goldreich}{1969}]{go69}
Goldreich P. \& Julian W.H., 1969, ApJ, 157, 869
\bibitem[\protect\citeauthoryear{Harding}{2008}]{ha08}
Harding A.K., Sterm J.V., Dyks J. \& Frackowiak M., 2008, ApJ, 608, 1378
\bibitem[\protect\citeauthoryear{Hirotani}{2008}]{hi08}
Hirotani K., 2008, ApJ, 688L, 25
\bibitem[\protect\citeauthoryear{Hirotani}{2006}]{hi06}
Hirotani K., 2006, ApJ, 652, 1475
\bibitem[\protect\citeauthoryear{Romani}{2010}]{ro10}
Romani R.W. \& Watters, K.P., 2010, ApJ, 714, 810
\bibitem[\protect\citeauthoryear{Muslimov}{2003}]{ro03}
Muslimov A.G. \& Harding A.K., 2003, ApJ, 588, 430
\bibitem[\protect\citeauthoryear{Ruderman}{1991}]{ru91}
Ruderman M.A., 1991, ApJ, 366, 261
\bibitem[\protect\citeauthoryear{Ruderman}{1975}]{ru75}
Ruderman M.A. \& Sutherland P.G., 1975, ApJ, 196, 51
\bibitem[\protect\citeauthoryear{Takata}{2006}]{ta06}
Takata J., Shibata S.,  Hirotani K. \& Chang H.-K., 2006, MNRAS, 366, 1310
\bibitem[\protect\citeauthoryear{Takata}{2007}]{ta07}
Takata J. \& Chang H.-K., 2007, ApJ, 670, 677
\bibitem[\protect\citeauthoryear{Takata}{2010}]{ta10}
Takata J, Wang Y \& Cheng K.S., 2010, ApJ, 715, 1318
\bibitem[\protect\citeauthoryear{Thompson}{2004}]{th04}
Thompson D.J., 2004, in Cheng K.S., Romero G.E., eds, Cosmic Gamma Ray
Sources. Dordrecht, Kluwer, p. 149
\bibitem[\protect\citeauthoryear{Trep}{2010}]{tr10}
Trep L., Hui C.Y., Cheng K.S., Takata J., Wang Y., Liu Z.Y. \& Wang N., 2010, MNRAS, in press
\bibitem[\protect\citeauthoryear{Venter}{2009}]{ve09}
Venter C., Harding A.K. \& Guillemot L., 2009, ApJ, 707, 800
\bibitem[\protect\citeauthoryear{Zhang}{1997}]{za97}
Zhang L. \& Cheng K.S., ApJ, 1997, 487, 370

\end{thebibliography}
\end{document}